\input psfig.tex
\documentstyle[aaspp]{article}

\def\lta{{\>\rlap{\raise2pt\hbox{$<$}}\lower3pt\hbox{$\sim$}\>}}
\def\gta{{\>\rlap{\raise2pt\hbox{$>$}}\lower3pt\hbox{$\sim$}\>}}

\lefthead{Spaans \& Carollo}
\righthead{Effective Star Formation Rates}

\begin{document}

\title{Effective Star Formation Rates for Cosmological Applications}

\author{Marco Spaans\footnote{Hubble Fellow}}

\affil{Harvard-Smithsonian Center for Astrophysics, 60 Garden Street, Cambridge, MA 02138}

\author{C. Marcella Carollo$^1$}

\affil{Department of Physics \& Astronomy, The Johns Hopkins University, Baltimore, MD 21218}

\begin{abstract}

Effective star formation rates in tabular form are computed which yield a
prescription for the star formation activity in model galaxies
as a function of ambient
density, metallicity, and stellar feedback. The effects of supernova
explosions on the thermal balance of the Interstellar Medium (ISM) and
the presence of a multi-phase ISM are explicitly included. The
resulting grid of models can be implemented easily in N-body codes for the
computation of star formation processes in merging galaxies and
cosmological simulations.

\end{abstract}

{\it subject headings}: galaxies: star formation - galaxies:
structure - ISM - molecular processes

\section{Introduction}

The past decade has seen tremendous advances in the field of numerical
cosmology (Navarro, Frenk \& White 1996 and references therein).
N-body simulations and hydro codes have benefitted from the large
increase in CPU and memory capabilities of the current generation of
supercomputers, and have reached a high enough sophistication to
describe the development of large scale structure in the universe and
the formation of galaxies with redshift. An important issue which
ultimately needs to be addressed, in order to compare theory with
observations of high redshift galaxies, is the formation of stars in
proto-galactic structures (Norman \& Spaans 1997, hereafter NS97, and
references therein).  Many parameterizations of the star formation
rate in the ambient Interstellar Medium (ISM) exist in the literature
and mostly derive from a Schmidt law applied to a sufficiently large
body of gas (Theis et al.~1992; Kauffmann \& White 1993; Spaans \&
Norman 1997, hereafter SN97).

Star formation is a local phenomenon which must find its
explanation in the stability and fragmentation of dense molecular
clouds.  Studies in our own Galaxy have focussed on the structure of
dense proto-stellar cores, along with the chemical and thermal balance of
star-forming regions (Helmich 1996, and reference therein).
These studies lend indirect support to a Schmidt
law, but emphasize the need to include explicitly the structure of the
multi-phase ISM to model accurately the most important heating and
cooling processes. A large unknown in these investigations is the relative
importance of feedback.
Supernova explosions and stellar radiation associated
with the process of star formation influence the global physical
structure of the interstellar gas which supports this process.

Ideally, in the cosmological context one would like to
solve for the properties of the thermal and chemical balance of the
ISM and the star formation rate simultaneously with the solution of
the gravitational N-body and hydrodynamical problem. Such an approach
is beyond what is currently feasible and one is forced to decouple the
ISM and star formation problem from the dynamical one. Along these
lines, the purpose of this work is to present a set of numerical
simulations which span a large range in density and metallicity,
include a quantitative prescription for feedback, and yield the
resulting star formation rate in the ambient medium in tabular form.
This work extends the methods developed in a
series of papers where the structure of the ISM and the importance of
feedback effects have been investigated for proto-galactic disks
(NS97), dwarf galaxies (SN97), massive proto-spheroidals in the Hubble
Deep Field (Spaans \& Carollo 1997), and elliptical galaxies (Carollo
\& Spaans 1998).

The strength of the procedure used in this work to produce the effective
star formation rates is its detailed treatment of molecular line
cooling, chemistry, and thermal balance (section 2) for a wide range of
physical conditions.  This allows for a detailed treatment of the
multi-phase structure of the ambient ISM, and the feedback of
stellar photons and supernova ejecta on its physical balance.
The star formation rate is computed in our approach through a Schmidt law
applied to the molecular phase of the ISM. The use of this empirical
law presumes that star formation as observed in our local environment is
representative of star formation in general. This seems a reasonable
approach for spiral galaxies (Kennicutt 1989).

In the following sections, the input physics are discussed.
The processes involved are complex, and various physical phenomena
influence the final outcome of the computation.
Therefore, one should keep the following major
calculational steps in mind. 1) The thermal and chemical balance of the
medium is determined locally and used to compute the amount of molecular
gas. 2) This allows a calculation of the formation rate of stars according to
a Schmidt law and
an initial mass function (IMF). 3) The produced radiation
is propagated across the computational grid in a radiative transfer
calculation, and used
to determine heating, ionization and dissociation rates. 4) The
supernova blast waves, produced as the end products of stellar evolution,
are propagated across the grid as well and yield
the local input of kinetic (gas bulk motions) and thermal (gas heating) energy,
as well as metals.
5) Steps 3 and 4 provide input for step 1. This entire procedure is followed
for many different ambient densities, to derive the enrichment and physical
state of the gas together with the corresponding star formation rate.

\section{The Model: Stars Embedded in a Multi-Phase ISM}

The models include three stellar and three gaseous components. The gas
phases include the cold molecular clouds, the warm neutral/ionized
medium, and the hot tenuous interiors of supernova bubbles. The phases
are assumed to be in pressure equilibrium and their chemical and
thermal balance is computed explicitly.
The theoretical background for the evolution of the multi-phase ISM in
primordial galactic structures is described in NS97 and SN97, where
the employed numerical methods and a discussion of the spatial and
temporal resolution of the adopted grid can also be found.
Although no dynamics are included explicitly, there is mass exchange
between the cold molecular and hot tenuous phase due to cloud
evaporation. The density dependence for mass evaporation is given by
$\propto E^{6/5}n_{\rm h}^{-4/5}$, with $E$ the supernova energy and
$n_{\rm h}$ the density of the hot phase (McKee \& Ostriker 1977).
In the following we
adopt for a grid cell a typical size of 20 pc, and time steps of 30-60 Myr.
Even though we determine stationary solutions, it is necessary to integrate
in time to determine the response of the ambient gas to the star formation
process, i.e.\ stellar evolution, propagation of supernova shock waves etc.

The stellar components are
divided according to their final evolutionary stages into massive
stars of more than 11 M$_\odot$, which explode as Type II supernovae,
and low-mass stars, which are assumed to loose their material in a
planetary nebula phase. The low-mass stars will become 0.6 M$_\odot$
white dwarfs (Weidemann \& Koester 1983). Stars with masses below 0.6
M$_\odot$ do not evolve during the lifetime of a galaxy. The third
class of stars comprises the stellar remnants in the form of white
dwarfs, neutron stars, and black holes.  A fraction of the white
dwarfs gives rise to Type Ia supernovae in the merging CO-dwarf
picture (see below). Below we describe in detail the set of equations
which are solved
for the star formation (SF) and feedback processes: chemical and
thermal balance, the occurrence and propagation of supernovae, metal
yields and enrichment, radiative transfer, and stellar life cycles.

\subsection{Star Formation}

Following Larson (1991) the star formation rate is calculated with a
Schmidt (1959) law applied to some volume in a galaxy with a gas mass $M$
$${{\partial M_{\rm cm}}\over{\partial t}}=
\alpha_{\rm SF} ({{n_{\rm cm}}\over{n_{\rm cm}^0}})^b,\eqno(1)$$
where the label ``cm'' indicates the cold molecular phase, $n_{\rm cm}$ is
a number density, $n_{\rm cm}^0=40$ cm$^{-3}$
and $b=1-2$. The coefficient $\alpha_{\rm SF}$ is normalized to the SF
rate as observed in molecular gas in the Solar vicinity
$$\alpha_{\rm SF}={{M}\over{M_{\rm g}}}\qquad {\rm M}_\odot
{\rm yr}^{-1},\eqno(2)$$
with $M$ normalized to the total body of gas $M_{\rm g}$ in the Milky Way
(see below). For our own Galaxy as a whole
this corresponds to an instantaneous rate of $\sim 3$
M$_\odot$ yr$^{-1}$ (Wyse 1997, private communication).
The generic value of $b$ is 1.3, but $b=2$
is adopted in those low metallicity regions where the
ambipolar diffusion time is shorter than the free-fall time of a
molecular cloud (see NS97 for a detailed discussion).

The density of the cold molecular phase follows from application of
the Field, Goldsmith, and Habing (1969) formalism to the thermal
balance of a multi-phase ISM.  The angular momentum of star-forming
clouds is not included beyond the accuracy of the local Schmidt law
applied to the cold molecular component. As such, we cannot address
the (important) question of cloud fragmentation, and the IMF is an input
parameter. We have assumed a time-independent Salpeter IMF $\phi
(m)\propto m^{-2.35}$. Stellar masses between 0.1 M$_\odot$ and 40
M$_\odot$ are considered. It follows that approximately 10\% of the
newly formed stellar mass is incorporated into massive stars with a
lifetime of no more $2\times 10^7$ years.

\subsection{Chemical Evolution}

The evolutionary end stages of stars produce metals which provide most
of the cooling (atomic and molecular lines) and dust opacity in a model galaxy.
It is thus vital to know the production and spatial distribution of metallic
species.

\subsubsection{Elemental Abundances in the Gas Phase}

The chemical evolution of elemental species has been discussed
extensively in the literature, and the equations as described in
Timmes, Woosley \& Weaver (1995) and Tantalo et al.~(1996) are
adopted. The main difference with our implementation is the explicit
three-dimensional spatial dependence of the chemical quantities.
The treatment of physical processes
is explicitly three-dimensional, and the spatial
enrichment of the model galaxy through the expansion fronts of
multiple supernovae is
included. While the supernova bubble expands with time, the rate at
which metals condense out into the cold molecular phase is computed to
determine the local enrichment. The cooling rate of the enriched gas
then determines whether a molecular phase can be supported, and at
what interstellar gas density the SF process takes place.  This
eliminates the assumption of a one zone system. Allthough there is no
infall of material, and the model galaxy is therefore a closed box, there is
mass exchange (in particular metals) between the grid cells through the action
of the SN expansion fronts.

For the stellar yields, three distinct components are specified: one
for the massive stars ($11M_\odot\le M\le 40M_\odot$) which become
Type II supernovae, one for the intermediate to low mass stars ($M\le
11M_\odot$) which become planetary nebulae, and one for the remnants
of the intermediate to low mass stars that become Type Ia supernovae
(Timmes et al.~1995).  The Type II supernova yields of Woosley
\& Weaver (1995) are adopted throughout. The yields from intermediate
to low mass stars were taken from the CNO models with $1M_\odot\le
M\le 8M_\odot$ of Renzini \& Voli (1981). We have adopted their case B
table for $Z=0.004Z_\odot$ and $Z=Z_\odot$, and applied linear
interpolation for the intermediate metallicities. For metallicities larger
than Solar we extrapolated the yields linearly. This extrapolation does not
strongly influence our results. By the time the ISM has been enriched to
super-Solar metalliities, most of the baryons in a model galaxy have been
incorporated into stars. The W7 model of
Nomoto, Thielemann \& Yokoi (1984) was adopted for the Type Ia
supernovae.

The average mass $\xi_i(p,t)$ and abundance $X_i(p,t)$ of the
elemental species in a grid cell centered on the point $p=(x,y,z)$ at
time $t$ are governed by
$${{d\xi_i(p,t)}\over{dt}}=-X_i(p,t)S(p,t)+\int_{M_{\rm min}}^{m_{\rm B}}
S(p,t-t_M)Y_{M,i}(t-t_M)I(M)dM+$$
$$\qquad C\int_{m_B}^{M_B}I(M)[\int_{\mu_{\rm min}}^{0.5}f(\mu )
S(p,t-t_{M_2})Y_{M,i}(t-t_{M_2})d\mu ]dM+$$
$$\qquad(1-C)\int_{m_B}^{M_B}S(p,t-t_M)Y_{M,i}(t-t_M)I(M)dM+$$
$$\qquad\int_{M_B}^{M_{\rm
max}}S(p,t-t_M)Y_{M,i}(t-t_M)I(M)dM.\eqno(3)$$ In the equation, $S(p,t)$
is the normalized SF rate, $Y_{M,i}$ are the elemental yields of
elements $i$ from stars of mass $M$, $I(M)$ is the initial mass
function whose lower and upper limits are $M_{\rm min}$ and $M_{\rm
max}$, $t_M$ is the lifetime of a star of mass $M$, and $t_{M_2}$ is the
lifetime of the secondary in the binary, which determines the time
required for the system to undergo a Type Ia event. In the models
this lifetime depends on the initial chemical composition as tabulated
in Bertelli et al.~(1994). The various integrals which appear in
Equation (3) represent the contributions of Type II and Type Ia
supernovae in the formulation of Matteucci \& Greggio
(1986). Specifically, the second integral reflects those binary
systems which have the necessary properties to produce Type Ia
supernovae, $m_B$ and $M_B$ are the lower and upper mass limit for the
total mass of the binary system, $f(\mu )$ is the distribution
function of their mass ratios, and $\mu_{\rm min}$ denotes the minimum
value of this ratio.  The lower integration limit is equal to 3
M$_\odot$, to ensure that the accreting white dwarf can reach the
Chandrasekhar limit. The upper mass limit is equal to 16 M$_\odot$ for
a binary system, based on the assumption that the maximum mass of the
primary that produces a carbon-oxygen white dwarf is 8 M$_\odot$.  The
constant $C$ indicates which fraction of all binary systems undergo
supernova events.  This constant is chosen equal to $C=0.007$ to reproduce the
solar $^{56}$Fe abundance (Timmes et al.~1995).

\subsubsection{Atomic and Molecular Species}

If one knows the local elemental abundances and the radiation field (discussed
below) then it is possible to determine the chemical composition of the
ambient gas.
To derive the abundances of atomic and molecular species, the
solution vector of a chemical network in point $p$ is computed according to
$$H(T(p),n_i(p))={{dn_i(p)}\over{dt}}=0.\eqno(4)$$ Here $H$ denotes
the chemical network, $T$ is the ambient temperature, and the chemical
abundances by number $n_i$ are determined in steady state. That is, the
evolution time of a galaxy, a free-fall time $\sim 500$ Myr, is assumed to
be long compared to the time to reach chemical equilibrium, $\sim$
50 Myr for diffuse H$_2$ gas and much shorter for all other molecular species.
Only in a very low metallicity environment, $Z\sim 0.005Z_\odot$, is the
formation time scale of H$_2$ as long as 500 Myr and intimately linked to
the dynamical evolution of the system.

The adopted reaction network is based on that used by van Dishoeck \&
Black (1986, 1989) to model the chemistry of translucent clouds,
combined with the sulfur chemistry as described in Drdla et
al.~(1989). The network has been checked against the UMIST database
(Millar et al.~1991), and only minor differences have been found. The
chemical network includes 24 elements, and isotopes of H, C and O. The
inclusion of many metals is important because they dominate the
ionization balance and differences in ionization potential become
important for extinctions of more than 2 mag. The network includes 1549
reactions among 215 species. We adopt the line and continuum
photo-dissociation cross sections based on the literature summarized
by van Dishoeck (1988).

\subsection{Thermal Balance}

The heating of the ISM involves radiative sources like stars as well as
mechanical sources like stellar winds and supernova expansion fronts.
Cooling is provided by line emission from hydrogen and metallic ions as
well as continuum processes. In order to determine the thermal state of the
gas, one needs to find the solutions ($\rho$,$T$) of the
fundamental energy ($E$) equation
$${{dE(T,\rho ,t)}\over{dt}}=\Gamma (T,\rho ,t)-\Lambda (T,\rho
,t).\eqno(5)$$
In this equation, $T$ is the temperature, $\rho$ is the gas density,
$\Gamma$ is the total heating rate and $\Lambda$ the total cooling
rate, both per unit of mass.
In the following subsections the various terms which
contribute to the overall balance are discussed.

\subsubsection{Heating by Supernovae}

The hydrodynamics of interstellar gas is not
treated explicitly in our work,
and the results of various numerical studies are
used to implement the contribution from mechanical processes such as
supernova blast waves.
We adopt the formulations of Bertelli et al.~(1994) for Type II, and of
Greggio \& Renzini (1983) for Type Ia supernovae. The total energy injection
due to supernovae of Type $X$ which last for a time $\Delta t$ is given by
$$\Gamma_X=\int_{\Delta t}\epsilon_X(t-t')R_X(t')dt',\eqno(6)$$ where
$R_X(t)$ is the number of supernovae per unit mass and time, and
$\epsilon_X(t)$ describes the thermalization of the supernova energy
as a function of time when cooling effects are properly taken into
account. For $\epsilon_X$ we adopt the results of Gibson (1995)
and Thornton et al.~(1997). From the latter authors we adopt the
radial evolution of the supernova remnants to determine the
distribution of metal-rich tenuous gas as it condenses out into the
cold molecular phase. The rate $R_X$ is computed from the local star
formation rate in grid point $p$ at time $t$ and depends on the assumed IMF.
We have also included in Equation (5) the work performed by supernova shocks
through the acceleration of molecular clouds. From Thornton et al.\ (1997) we
take the fraction of the initial supernova explosion energy which ends up as
kinetic energy of the ISM.

\subsubsection{Heating by Stellar Winds}

The rate of energy injection by stellar winds has a similar form, namely:
$$\Gamma_W=\int_{\Delta t}\epsilon_W(t-t')R_W(t')dt',\eqno(7)$$ where
$R_W(t)$ is the number of stars per unit mass and time expelling their
envelopes during the time interval $\Delta t$. The quantity
$\epsilon_W$ has the same meaning as $\epsilon_X$. A star which is
shedding mass, deposits an energy of (Gibson 1994)
$$\epsilon_{W0}=0.15M_{ej}(M)(Z/Z_\odot )^{0.75}v(M)^2,\eqno(8)$$
where $M_{ej}$ is the amount of mass ejected by a star of mass $M$ and
$v(M)$ is the velocity of the ejected material. The velocity $v(M)$ is
assumed to be equal to the maximum of the terminal velocity of the
wind $\approx 20$ km s$^{-1}$ or the galaxy velocity dispersion.  The
adopted process of thermalization is as described in Bertelli et
al.~(1994). The magnitude of this heating source is not that
accurately known, but it does not dominate the thermal balance of the
ISM.

\subsubsection{Heating by Ultraviolet Photons}

The ionization heating and charge balance of the medium is explicitly
included through the Saha equation applied to H, He, C, O, N, S, Fe,
Si, and Mg.  The energy density and shape of the radiation field
follows from
$$E_{\rm UV}=\int_{\Delta t}\epsilon_{\rm UV}(t-t')R_{\rm UV}(t')dt',\eqno(9)$$
where all the symbols have their usual meaning. In the determination of
$\epsilon_{\rm UV}$ we use the stellar population models of Leitherer
et al.\ (1996) for a Salpeter IMF.

To assess which fraction of the ultraviolet energy is available for
heating, ionization and dissociation,
a continuum radiative transfer calculation is
required.  In the literature (Granato et al.~1997 and references
therein) one finds a typical number of 0.01 for the fraction of the
ultraviolet light which is not absorbed by dust and re-radiated as
infrared radiation. Because the abundances of metals and therefore of
dust may vary strongly across the galaxy, the Monte Carlo radiative
transfer method of Spaans (1996) is adopted to compute the strength of the
average radiation field from the SF rate and the local metallicity
(SN97). The star formation rate used in the radiative transfer computation
is a local average over a grid cell, and yields a local source function
for the ultraviolet radiation field. This way local fluctuations in the
radiation field strength are taken into account in the chemical and
thermal balance.

For the molecular phase with metallicities larger then 1\% of Solar,
the dominant heating mechanism is the emission of
photo-electrons from dust grains. The formalism of Bakes \& Tielens (1994)
is adopted with the implementation as described in Spaans et al.~(1994).
Part of the photo-electric heating of the molecular ISM is provided by large
molecules like PAHs (polycyclic aromatic hydrocarbons), and their abundance is
assumed to be equal to 10\% of the gas phase carbon.

\subsubsection{Heating by Proto-Galactic Collapse}

Since the purpose of the effective SF rates is to study the gas phase processes
during the formation of
galaxies, we also include the heating which results from the infall of
material on a time scale $\tau_{\rm ff}$ and leads to the formation of
the galaxy (Binney \& Tremaine 1987). This is an aspect
of our calculations where a coupling with realistic gas dynamics from
a cosmological formalism would be most desireable.
Decoupling the ISM from the dynamics, we adopt a simplified approach here.
If $R_i$ is the initial radius of the proto-galaxy, then
$$\tau_{\rm ff}=\pi\surd{{{R_i^3}\over{2GM}}},\eqno(10)$$
where $M$ is the total mass of the galaxy. If no other energy sources are
present, a proto-galactic cloud at the virial temperature cools on a time scale
$$\tau_{\rm cool}=6.3\times 10^5(R_i/10{\rm kpc})^2\qquad {\rm
yr}.\eqno(11)$$ At the end of the collapse phase the system is assumed
to have an energy given by
$$E_{\rm vir}=E_{\rm vir,0}(1-{{\tau_{\rm cool}}
\over{\tau_{\rm ff}}}),\eqno(12)$$
where $E_{\rm vir,0}$ is the initial virial energy.
The remaining energy at the end of the collapse phase
is converted into heat at the rate
$$\Gamma_{\rm c}={{E_{\rm vir}}\over{\tau_{\rm ff}}}.\eqno(13)$$ This
source of heating remains active as long as the age of the galaxy is
less than $\tau_{\rm ff}$, and sets the initial gas temperature. The latter
is very important for the early low metallicity chemical balance which pertains
in the primordial gas.
We adopt a collapse factor $\lambda\approx 30$ (NS97), i.e.~$R_i\sim 300$ kpc.

\subsubsection{Relative Importance of Heating Processes and Galactic Winds}

As mentioned above, the heating of the molecular gas is dominated by
photo-electric emission for metallicities in excess of $0.01Z_\odot$.
It is this rate which determines to a large extent the thermal and
chemical balance of the multi-phase ISM, and therefore the effective
star formation rate supported by the ambient molecular gas. Nevertheless,
the other heating terms are of importance as well during certain epochs.
The proto-galactic collapse phase provides the initial thermal structure
of the ambient gas prior to the formation of stars. It thus fixes the chemical
abundances of the primordial galaxy.
During periods of massive star formation, the energy input from supernovae
becomes comparable or even larger than the photo-electric heating rate
in small regions of the galaxy where the supernova blast wave interacts
with the ambient interstellar gas.

If the supernova energy input dominates during some period of time, the
temperature or kinetic energy of the interstellar gas can increase and
may become comparable to the gravitational binding energy of the model galaxy.
In this case
a wind develops which is assumed to quench the SF process (e.g. Larson 1974;
Carlberg 1984; Arimoto \& Yoshii 1987; Matteucci \&
Tornamb\'e 1987; De Young \& Heckman 1991).

\subsection{Cooling by Line and Continuum Processes}

The cooling rate depends crucially on the abundance of certain atoms
and molecules like C, O, H$_2$ and CO, and the ability for radiation
to escape from the medium, i.e.~radiative transfer. For temperatures
above $10^4$ K, the metallicity dependent tabulations of Sutherland \&
Dopita (1993) and the cooling curves of Ferrara (1996, private
communication) are adopted.  For these high temperatures the cooling is
dominated by thermal bremsstrahlung and by atomic hydrogen around $10^4$ K.
Unlike other treatments
(Chiosi et al.~1997) we do not use cooling functions below
$10^4$ K, but solve the statistical equilibrium and the radiative
transfer problem explicity (Spaans \& van Dishoeck 1997; SN97).

The fine-structure lines of C$^+$, C, O, Si$^+$, Si, S$^+$, S, Fe$^+$,
and Fe are included, as well as the rotational lines of CO and
H$_2$. The vibrational excitation of the latter molecule is also
included.  The thermal balance of the molecular phase strongly depends
on the chemistry and thereby influences the effective SF rate of the
interstellar gas. A detailed treatment of optical depth effects in the
cooling lines is therefore required to
determine the various densities and temperatures of the components of
the multi-phase ISM (NS97).
Note that in redshift dependent calculations, the temperature of the Cosmic
Microwave Background (CMB) should be taken explicitly into account for
redshifts larger than 5.  The
latter effect is of particular importance for the excitation of CO
molecules, and influences the effective cooling rate of
molecular gas (Silk \& Spaans 1997). Still, we do not expect the effective
star formation rate to change much and the application of our results is
likely to be to lower redshift objects. Therefore we have considered only
a 3 K CMB.

\section{Results and Discussion}

\subsection{Model Characteristics}

In the computation of the star formation rates,
the total body of gas simulated is $M_{\rm
g}=10^{10}$ M$_\odot$, i.e.~the gas mass of the Milky Way. The effective
star formation rates presented here
should be scaled linearly when implemented in dynamical
simulations with some mass resolution element $M_{\rm res}$.
To model the effect of outflows on the star formation
process the body of gas is assumed to be embedded in a $M_{\rm
pot}=5\times 10^{11}$ M$_\odot$ spherical potential with a $r^{-2}$
density profile outside a constant density core (Spaans \& Carollo
1997).  The energy estimates of SN97 are used to determine when the
gas becomes unbound and star formation is quenched. We have assumed a
generic SN kinetic energy of $10^{51}$ ergs, and that 10\% of the
initial SN energy ends up as turbulent kinetic energy of the
interstellar clouds. The amount of kinetic energy transferred from the
SN expansion front to the ISM is not that well
determined. Nevertheless, recent one-dimensional numerical simulations
by Thornton et al.~(1997) indicate that 10\% is a reasonable
value. The value of $M_{\rm pot}$ determines the binding energy of the gas
in a linear fashion. The maximum possible
star formation rate, i.e.\ when a galactic
wind does not yet develop, therefore scales with $M_{\rm pot}$ in
our formalism. This scaling does not take into account the nature of the
outflow which develops for a galaxy with a specific density profile.
If the potential is not static the local density changes should
be well captured by our formalism (barring shocks). The maximum star formation
rate on the other hand would change, and our method does not accurately
capture these effects.

The characteristic time scale and absolute rate of star formation depend on the
thermal and chemical balance of the ISM. For the numerical simulations
presented here, a grid was set up in parameter space which includes
total hydrogen number densities $n_{\rm H}=10^{-2}-10^6$ cm$^{-3}$ and
metallicities $Z=5\times 10^{-3}-3$ $Z_\odot$. The locally produced
stellar radiation field and the input of kinetic and thermal energy by
supernova explosions render our equation for the local star
formation rate non-linear
due to feedback effects. The numerical simulations have therefore been
performed for a range of ``background'' SF rates $S_{\rm b}=0.3-300$
M$_\odot$ yr$^{-1}$. The SF rates at the low end of the background
range can be much larger than 300 M$_\odot$ yr$^{-1}$ initially, but
as $S$ increases, feedback effects in the form of outflows limit the
sustained SF rate to $\sim 10^3$ M$_\odot$ yr$^{-1}$. This result
depends on the adopted values of $M_{\rm g}$ and $M_{\rm pot}$ as discussed
above. In practical applications, the background star formation rate should
be defined be neighboring grid cells or a previous time step.

\subsection{Physical Processes}

Before turning to the specific results, it is good to bear in mind the
following physical processes which determine to a large extent the
conclusions of this paper. For metallicities of less than a percent of
Solar, the chemistry is dominated by H, He, Li, their ions, H$^-$, and
molecules
like H$_2$ and HD. The latter two, together with atomic hydrogen, provide
the cooling of the ambient gas. The dominant formation route of H$_2$
involves the H$^-$ radical rather than dust grains. A consequence, of
importance for the results presented here, is that star formation is
a global process. That is, the lack of dust allows stellar UV photons to
traverse throughout the medium almost unattenuated and to
influence the physical
structure of the ISM over very large distances. In NS97 it is shown that
this mode of star formation is self-regulating due to the strong dependence
of the H$^-$ abundance on the radiation field strength shortward of 912 \AA,
and the fact that H$_2$ is only dissociated by photons longward of 912 \AA.
The typical H$_2$ driven
star formation rate for a proto-galactic disk is small and of
the order of 0.01-0.1 M$_\odot$ yr$^{-1}$. The relevance of this number
for low surface brightness galaxies will be discussed in a forthcoming paper
(Spaans, Mihos, \& McCaugh 1998; in preparation).

For metallicities in excess of a few percent of Solar, the above picture
changes dramatically. The presence of metallic atoms enhances the cooling
rate and allows the formation of molecules like OH, H$_2$O and CO. Also the
formation of H$_2$ proceeds much more efficiently on the surfaces of dust
grains. Due to the increased dust opacity, the formation of stars is a local
process. Stellar photons are absorbed close to the source and the presence
of metallic atoms and molecules allows a large fraction of this input energy
to be radiated away efficiently.

The following phases can then be distinguished in the evolution of the
ambient medium: Only a small fraction of baryonic matter is
converted into stars due to H$_2$ and HD cooling in a very low metallicity
medium. The regions where the H$_2$ abundance is large are
characterized by a kinetic temperature of $\sim 1000$ K, due to
vibrational line cooling, and an ambient pressure of $2\times 10^3$ K
cm$^{-3}$. The efficient formation of stars begins with a
transition to a multi-phase ISM at $5\times 10^3$ K cm$^{-3}$, which includes
cold molecular clouds.
This transition occurs at a metallicity of the order of $\sim$0.03$Z_\odot$.
The multi-phase ISM is able to cool away most of the UV stellar photons through
infrared lines and dust continuum emission.

\subsection{Model Results}

Figure 1 presents the general dependence on metallicity and hydrogen
number density
(for a mass $M_{\rm g}$) and a background star formation rate
$S_{\rm b}=3$ M$_\odot$ yr$^{-1}$. The dots in Figure 1 (and Figure 2)
indicate individual entries in Table 1. Note here that the model results
presented in Table 1 have been interpolated with a low order polynomial to
create a grid which is homogeneous in both density and metallicity.
Around the threshold metallicity of $\sim 0.03$ Z$_\odot$, the
sustained SF rate is a strong function of the ambient enrichment. Because
the cooling rate cannot keep up with the ambient heating, a small rise in
the SF rate raises the temperature of the ISM to above $\sim 1000$ K where
H$_2$ is collisionally destroyed. Conversely, a small increase in metallicity
increases both the cooling and the H$_2$ formation rate on dust grains, and
therefore the sustained SF rate. Once a multi-phase ISM has been
firmly established for metallicities roughly within an order of magnitude of
Solar, the star formation rate does not vary with metallicity by more than a
factor of two. This relative insensitivity to metallicity in a multi-phase
ISM is due to the ability of the gas to cool through infrared line transitions
and dust emission.

Note that the 0.01 and 0.02$Z_\odot$ curves are the same for
number densities below 0.03 cm$^{-3}$, but not the same as the 0.005 and
0.05$Z_\odot$ curves below that number density. Below a metallicity of
$\sim$0.01Z$_\odot$, a multi-phase ISM does not exist. The 0.005$Z_\odot$ curve
therefore pertains to a medium cooled predominantly by H, H$_2$ and HD.
Conversely, for metallicities close to the phase transition, but at low
densities, the pressure-density curve which determines the phase structure
of the ambient gas, does not vary much with metallicity. This is caused by the
fact that for these parameters the heating and cooling rates vary
proportionally.

Figures 1 and 2 also illustrate the importance of feedback to limit the
global SF rate to no more than $\sim 10^3$ M$_\odot$ yr$^{-1}$ for the
masses and ambient densities considered here. A system with ($M_{\rm
pot}=5\times 10^{11}$ $M_\odot$, $M_{\rm g}=10^{10}$ $M_\odot$)
starts to unbound its ISM kinematically for a star
formation rate of $\sim 10^3$ M$_\odot$ yr$^{-1}$, even though the
thermal energy contributed to the molecular ISM can be largely
radiated away through line and dust emission. This causes a break in
the star formation rate with density for a given metallicity. For low densities
and SF rates the slopes in Figures 1 and 2 are for a large part dictated by the
assumed Schmidt law. This limit to the star formation rate
should be indicative for the formation of $L^*$-type
galaxies, and appears to be in agreement with current results for the
Hubble Deep Field (Madau et al.\ 1996).
A dynamical implementation of our results will yield a
more accurate determination of these effects since they could account
for re-accretion and the extent of the dark matter potential.

It is good to take note of the
time scales to reach chemical equilibrium in media of varying metallicity.
For a low metallicity environment of 0.5 percent of Solar,
the time to reach
chemical equilibrium for the H$_2$ abundance in a cloud of 100 cm$^{-3}$ is
approximately 500 Myr for a background radiation field of 0.1 in units of
$J_{-21}$. In the multi-phase ISM the corresponding time scale is a bit less
than 20 Myr. It is the H$_2$ abundance which provides the initial cooling
and drives the ion-molecule chemistry
required for the formation of molecules like OH, CH, H$_2$O, CO etc.
The decrease in this time scale with metallicity therefore gives a direct
measure of the efficiency with which stars can be formed.

These time scales are also relevant to the blow-out phenomenon in
flattened galaxies. A supernova
shock front originating from the disk of a galaxy and travelling at 100 km
s$^{-1}$ will reach a height of 1 kpc in 10 Myr. This time scale is of the
same order as the chemical equilibrium time as well as the cooling time of
the gas. The hot metal-rich cavity gas therefore does not have enough time
to cool and form (molecular) clouds which are more robust and could help
to retain the metals, even in a multi-phase ISM (Spaans \& Carollo 1997).

\subsection{Sensitivity to Model Assumptions and Parameters}

Some of the numbers which go into the calculation have a large influence
on the final result and are also poorly determined. These will be discussed
below.

1) An optimal metallicity range of
$\sim 0.04-0.2Z_\odot$ seems to exist, for which SF can proceed more
efficiently at high densities (SN97).
This effect is due to the absence of magnetic
support for low metallicity molecular clouds. A low ionization fraction
yields an ambipolar diffusion time scale which is shorter than the
free-fall time of a molecular cloud. The calculations presented here limit
this increase in the SF rate
to about of factor of two due to the grain chemistry adopted at
high extinction. Deep inside molecular clouds grains are predominantly
neutral and provide an important sinc for the free electrons followed by
rapid recombination with (metallic) ions (Lepp \& Dalgarno 1988). This leads
to a medium with a higher fractional ionization compared to the case where
electron recombination dominates. The rates for these processes are
typically 10$^{-6}$ cm$^3$ s$^{-1}$, but are not accurately known due to
their dependence on the grain size distribution.
Additional calculations at a selected number of densities which
involve a chemical network with more detailed charge transfer effects
and exchange rates larger by a factor of three,
indicate that the star formation rates in the optimal metallicity
range may be larger by about 100\%. The more conservative estimates
are retained here because only cloud collapse models could provide a
definite answer. These seem to indicate that diffusion of the magnetic field
aids the formation of stars, but is not a sufficient condition by itself
(Gammie 1997, private communication).

2) Related to the issue of molecular clouds at high visual extinctions, is
the matter of grid resolution. The implemented method allows the gas to
be resolved down to the 20 pc level, i.e.\ individual molecular clouds.
Any increase in grid resolution is therefore not expected to change our
results. Still, the star formation process itself involves the fragmentation
of molecular cloud cores and smaller scales should therefore be probed to
put the adopted Schmidt star formation law on a firmer basis. Such an increase
in resolution must be accompanied by a hydrodynamic treatment of the gas to
accurately model for instance
the effects of shocks. The latter are very dissipative
phenomena and strongly influence the thermal balance of the ambient gas.
It is expected that in a high metallicity medium, molecules will
be (re-)formed efficiently
in the shock's wake. Therefore, the energy will be efficiently radiated away
in the form of long wavelength photons and the gas can condense to high
densities. This will increase the effective star formation rate, although
the magnitude of the effect lies beyond the scope of the present work.

3) The heating rate due to supernova explosions can become comparable to the
UV heating rate in some parts of the model galaxy. Nevertheless,
the former number is not accurately known
and we have therefore run models where the thermalized fraction of the
input supernova energy is a factor of 5 larger. For low metallicities in
a H$_2$ cooling dominated medium, the temperature of the gas rises
with a factor of a few, from $\sim$ 1000 to $\sim$ 4000 K. This increases
the collisional destruction rate of molecular hydrogen, the dominant coolant,
and therefore
inhibits further star formation. Once a multi-phase ISM has been established,
the ambient gas is capable of radiating away the extra input energy efficiently
and the results for the star formation rates do not change by more than
30-50\%.

4) We have ignored the contribution from Type Ib supernovae. It appears that
radiative stellar winds from massive stars may be significantly suppressed
at low metallicities, but that winds driven by binary star evolution could
increase the number of Type Ib supernovae in the early universe (Timmes 1998,
private communication). Such an effect could increase the relative abundance
of oxygen. Since atomic oxygen is an important coolant and strongly influences
the chemistry of OH and CO, our results potentially can be altered by a
higher Type Ib supernova rate. For a higher oxygen production rate of a factor
of three, it turns out that the cooling rate due to [OI] is not strongly
influenced because the main cooling line, [OI] 63 $\mu$m, is optically thick.
The abundance of CO is favorably increased and the molecular phase can cool
more efficiently. For the adopted factor of three, the star formation rate
maximally increases by 40\%.

If one combines the uncertainties 1-4, it appears
that the magnitude of the star formation rates presented here
can change by at most a factor of
four, and are typically accurate to a factor of two. We feel that the
existence of a threshold metallicity, as depicted in Figures 1 and 2, is quite
robust since it derives from a detailed treatment of the thermal balance of
the ISM for a range of background radiation fields.

\section{Conclusions}

We have derived star formation rates in tabular form
as functions of the ambient
density, metallicity, and stellar feedback in proto-galaxies with a
given total gas mass. We have explicitely
included the effects of supernova explosions on the thermal balance of
the ISM, and the transition to a multi-phase ISM.  The
strength of the method used to produce the effective star formation rates
is the detailed treatment of molecular line cooling, chemistry, and
thermal balance. The tabulated star formation rates can be implemented in
N-body codes for the computation of star formation processes in
merging galaxies and cosmological simulations. The accuracy of
the presented numbers is expected to be of the order of a factor of two, and
in any case the rates should not change by more than a factor of four.
The full set of numerical results is included in Table 1, and can be
obtained in machine readable form from MS.

Although we have included the effects of feedback self-consistently,
the present calculations cannot capture the effects of massive shocks
associated with a large collection of SN explosions propagating
through the ISM. A complete treatment of the latter requires the
incorporation of the numerical scheme adopted here in a full hydro
treatment. Investigations along these lines are being pursued.

\acknowledgments The authors are grateful to Colin Norman and Frank Timmes
for stimulating comments.
MS and CMC are supported by NASA through grants
HF-01101.01-97A and HF-1079.01-96a, respectively, awarded by the Space
Telescope Institute, which is operated by the Association of
Universities for Research in Astronomy, Inc., for NASA under contract
NAS 5-26555.

\newpage

\vspace*{-2.2truecm}
\begin{tabular}{c c c c c c c c c c}
\multicolumn{10}{c}{Table 1}\\
\multicolumn{10}{c}{Effective Star Formation Rates (SF) as Functions of Metallicity,}\\
\multicolumn{10}{c}{Background Star Formation Rate (BSF) and Density}\\
\hline
\hline
Z/Z$_\odot$$^b$&$n_{\rm H}$ (cm$^{-3}$)$^b$&SF$^a$&BSF$^a$&SF&BSF&SF&BSF&SF&BSF\\
\hline
5.00E-03& 1.00E-02&  0.01& 0.30& 0.01& 3.00& 0.01& 30.0& 0.01& 300.0\\
5.00E-03& 2.00E-02&  0.01& 0.30& 0.01& 3.00& 0.01& 30.0& 0.01& 300.0\\
5.00E-03& 5.00E-02&  0.03& 0.30& 0.01& 3.00& 0.01& 30.0& 0.01& 300.0\\
5.00E-03& 1.00E-01&  0.10& 0.30& 0.04& 3.00& 0.01& 30.0& 0.01& 300.0\\
5.00E-03& 2.00E-01&  0.13& 0.30& 0.10& 3.00& 0.01& 30.0& 0.01& 300.0\\
5.00E-03& 5.00E-01&  0.19& 0.30& 0.12& 3.00& 0.02& 30.0& 0.01& 300.0\\
5.00E-03& 1.00E-00&  0.40& 0.30& 0.21& 3.00& 0.06& 30.0& 0.01& 300.0\\
5.00E-03& 2.00E-00&  0.62& 0.30& 0.39& 3.00& 0.10& 30.0& 0.01& 300.0\\
5.00E-03& 5.00E-00&  0.81& 0.30& 0.58& 3.00& 0.13& 30.0& 0.01& 300.0\\
5.00E-03& 1.00E+01&  0.99& 0.30& 0.67& 3.00& 0.19& 30.0& 0.02& 300.0\\
5.00E-03& 2.00E+01&  1.10& 0.30& 0.84& 3.00& 0.31& 30.0& 0.03& 300.0\\
5.00E-03& 5.00E+01&  1.42& 0.30& 0.98& 3.00& 0.41& 30.0& 0.05& 300.0\\
5.00E-03& 1.00E+02&  1.74& 0.30& 1.19& 3.00& 0.55& 30.0& 0.09& 300.0\\
5.00E-03& 2.00E+02&  1.96& 0.30& 1.30& 3.00& 0.69& 30.0& 0.12& 300.0\\
5.00E-03& 5.00E+02&  2.11& 0.30& 1.49& 3.00& 0.81& 30.0& 0.19& 300.0\\
5.00E-03& 1.00E+03&  2.40& 0.30& 1.60& 3.00& 0.93& 30.0& 0.24& 300.0\\
5.00E-03& 2.00E+03&  2.61& 0.30& 1.84& 3.00& 1.08& 30.0& 0.30& 300.0\\
5.00E-03& 5.00E+03&  2.80& 0.30& 1.98& 3.00& 1.20& 30.0& 0.41& 300.0\\
5.00E-03& 1.00E+04&  2.93& 0.30& 2.12& 3.00& 1.39& 30.0& 0.51& 300.0\\
5.00E-03& 2.00E+04&  3.04& 0.30& 2.31& 3.00& 1.54& 30.0& 0.62& 300.0\\
5.00E-03& 5.00E+04&  3.12& 0.30& 2.43& 3.00& 1.67& 30.0& 0.73& 300.0\\
5.00E-03& 1.00E+05&  3.24& 0.30& 2.75& 3.00& 1.85& 30.0& 0.81& 300.0\\
5.00E-03& 2.00E+05&  3.36& 0.30& 2.89& 3.00& 1.98& 30.0& 0.90& 300.0\\
5.00E-03& 5.00E+05&  3.41& 0.30& 2.96& 3.00& 2.12& 30.0& 0.99& 300.0\\
5.00E-03& 1.00E+06&  3.50& 0.30& 3.03& 3.00& 2.23& 30.0& 1.08& 300.0\\
\hline
1.00E-02& 1.00E-02&  0.10& 0.30& 0.10& 3.00& 0.10& 30.0& 0.10& 300.0\\
1.00E-02& 2.00E-02&  0.10& 0.30& 0.10& 3.00& 0.10& 30.0& 0.10& 300.0\\
1.00E-02& 5.00E-02&  0.13& 0.30& 0.10& 3.00& 0.10& 30.0& 0.10& 300.0\\
1.00E-02& 1.00E-01&  0.19& 0.30& 0.12& 3.00& 0.10& 30.0& 0.10& 300.0\\
1.00E-02& 2.00E-01&  0.28& 0.30& 0.19& 3.00& 0.14& 30.0& 0.10& 300.0\\
1.00E-02& 5.00E-01&  0.40& 0.30& 0.29& 3.00& 0.20& 30.0& 0.12& 300.0\\
1.00E-02& 1.00E-00&  0.63& 0.30& 0.43& 3.00& 0.32& 30.0& 0.18& 300.0\\
1.00E-02& 2.00E-00&  0.75& 0.30& 0.58& 3.00& 0.43& 30.0& 0.26& 300.0\\
1.00E-02& 5.00E-00&  0.89& 0.30& 0.69& 3.00& 0.52& 30.0& 0.38& 300.0\\
1.00E-02& 1.00E+01&  1.06& 0.30& 0.81& 3.00& 0.64& 30.0& 0.51& 300.0\\
1.00E-02& 2.00E+01&  1.20& 0.30& 0.95& 3.00& 0.72& 30.0& 0.62& 300.0\\
1.00E-02& 5.00E+01&  1.43& 0.30& 1.08& 3.00& 0.84& 30.0& 0.72& 300.0\\
1.00E-02& 1.00E+02&  1.72& 0.30& 1.23& 3.00& 0.97& 30.0& 0.84& 300.0\\
1.00E-02& 2.00E+02&  1.98& 0.30& 1.45& 3.00& 1.12& 30.0& 0.97& 300.0\\
1.00E-02& 5.00E+02&  2.31& 0.30& 1.60& 3.00& 1.32& 30.0& 1.11& 300.0\\
1.00E-02& 1.00E+03&  2.53& 0.30& 1.82& 3.00& 1.49& 30.0& 1.28& 300.0\\
1.00E-02& 2.00E+03&  2.81& 0.30& 1.99& 3.00& 1.58& 30.0& 1.40& 300.0\\
1.00E-02& 5.00E+03&  3.02& 0.30& 2.29& 3.00& 1.72& 30.0& 1.58& 300.0\\
1.00E-02& 1.00E+04&  3.18& 0.30& 2.51& 3.00& 1.90& 30.0& 1.69& 300.0\\
1.00E-02& 2.00E+04&  3.39& 0.30& 2.78& 3.00& 2.06& 30.0& 1.80& 300.0\\
1.00E-02& 5.00E+04&  3.53& 0.30& 2.97& 3.00& 2.25& 30.0& 1.97& 300.0\\
1.00E-02& 1.00E+05&  3.86& 0.30& 3.14& 3.00& 2.46& 30.0& 2.13& 300.0\\
1.00E-02& 2.00E+05&  4.07& 0.30& 3.27& 3.00& 2.61& 30.0& 2.34& 300.0\\
1.00E-02& 5.00E+05&  4.21& 0.30& 3.43& 3.00& 2.87& 30.0& 2.49& 300.0\\
1.00E-02& 1.00E+06&  4.48& 0.30& 3.59& 3.00& 2.99& 30.0& 2.61& 300.0\\
\hline
\end{tabular}
\newpage
\begin{tabular}{c c c c c c c c c c}
\hline
2.00E-02& 1.00E-02&  0.10& 0.30& 0.10& 3.00& 0.10& 30.0& 0.10& 300.0\\
2.00E-02& 2.00E-02&  0.12& 0.30& 0.10& 3.00& 0.10& 30.0& 0.10& 300.0\\
2.00E-02& 5.00E-02&  0.18& 0.30& 0.15& 3.00& 0.10& 30.0& 0.10& 300.0\\
2.00E-02& 1.00E-01&  0.27& 0.30& 0.22& 3.00& 0.16& 30.0& 0.10& 300.0\\
2.00E-02& 2.00E-01&  0.62& 0.30& 0.51& 3.00& 0.27& 30.0& 0.18& 300.0\\
2.00E-02& 5.00E-01&  1.09& 0.30& 0.92& 3.00& 0.47& 30.0& 0.26& 300.0\\
2.00E-02& 1.00E-00&  1.74& 0.30& 1.34& 3.00& 0.75& 30.0& 0.41& 300.0\\
2.00E-02& 2.00E-00&  2.96& 0.30& 2.17& 3.00& 1.08& 30.0& 0.62& 300.0\\
2.00E-02& 5.00E-00&  3.89& 0.30& 3.19& 3.00& 1.68& 30.0& 0.89& 300.0\\
2.00E-02& 1.00E+01&  5.54& 0.30& 4.89& 3.00& 2.76& 30.0& 1.24& 300.0\\
2.00E-02& 2.00E+01&  6.79& 0.30& 6.08& 3.00& 3.92& 30.0& 2.05& 300.0\\
2.00E-02& 5.00E+01&  8.73& 0.30& 7.83& 3.00& 5.13& 30.0& 3.16& 300.0\\
2.00E-02& 1.00E+02&  9.89& 0.30& 8.83& 3.00& 6.25& 30.0& 4.27& 300.0\\
2.00E-02& 2.00E+02&  11.6& 0.30& 10.1& 3.00& 7.93& 30.0& 5.16& 300.0\\
2.00E-02& 5.00E+02&  13.8& 0.30& 12.2& 3.00& 9.05& 30.0& 7.49& 300.0\\
2.00E-02& 1.00E+03&  16.1& 0.30& 14.9& 3.00& 10.6& 30.0& 8.99& 300.0\\
2.00E-02& 2.00E+03&  18.4& 0.30& 16.8& 3.00& 11.8& 30.0& 9.79& 300.0\\
2.00E-02& 5.00E+03&  21.6& 0.30& 19.6& 3.00& 13.9& 30.0& 10.8& 300.0\\
2.00E-02& 1.00E+04&  24.9& 0.30& 22.0& 3.00& 16.3& 30.0& 12.9& 300.0\\
2.00E-02& 2.00E+04&  28.5& 0.30& 24.8& 3.00& 19.5& 30.0& 14.5& 300.0\\
2.00E-02& 5.00E+04&  33.6& 0.30& 27.2& 3.00& 22.6& 30.0& 17.3& 300.0\\
2.00E-02& 1.00E+05&  39.4& 0.30& 31.5& 3.00& 26.8& 30.0& 21.8& 300.0\\
2.00E-02& 2.00E+05&  46.4& 0.30& 37.4& 3.00& 30.4& 30.0& 24.9& 300.0\\
2.00E-02& 5.00E+05&  54.9& 0.30& 41.3& 3.00& 33.9& 30.0& 27.4& 300.0\\
2.00E-02& 1.00E+06&  61.4& 0.30& 46.1& 3.00& 37.7& 30.0& 30.6& 300.0\\
\hline
4.00E-02& 1.00E-02&  0.20& 0.30& 0.13& 3.00& 0.12& 30.0& 0.12& 300.0\\
4.00E-02& 2.00E-02&  0.31& 0.30& 0.24& 3.00& 0.21& 30.0& 0.20& 300.0\\
4.00E-02& 5.00E-02&  0.45& 0.30& 0.39& 3.00& 0.35& 30.0& 0.34& 300.0\\
4.00E-02& 1.00E-01&  0.68& 0.30& 0.58& 3.00& 0.49& 30.0& 0.47& 300.0\\
4.00E-02& 2.00E-01&  1.34& 0.30& 0.97& 3.00& 0.88& 30.0& 0.85& 300.0\\
4.00E-02& 5.00E-01&  2.01& 0.30& 1.68& 3.00& 1.43& 30.0& 1.37& 300.0\\
4.00E-02& 1.00E-00&  4.89& 0.30& 3.29& 3.00& 2.94& 30.0& 2.81& 300.0\\
4.00E-02& 2.00E-00&  7.58& 0.30& 5.74& 3.00& 4.89& 30.0& 4.67& 300.0\\
4.00E-02& 5.00E-00&  14.6& 0.30& 10.3& 3.00& 8.43& 30.0& 8.19& 300.0\\
4.00E-02& 1.00E+01&  22.7& 0.30& 18.4& 3.00& 16.2& 30.0& 15.6& 300.0\\
4.00E-02& 2.00E+01&  41.8& 0.30& 34.2& 3.00& 29.8& 30.0& 27.3& 300.0\\
4.00E-02& 5.00E+01&  89.6& 0.30& 79.2& 3.00& 71.4& 30.0& 66.7& 300.0\\
4.00E-02& 1.00E+02&  126& 0.30& 113& 3.00& 99.2& 30.0& 92.3& 300.0\\
4.00E-02& 2.00E+02&  165& 0.30& 141& 3.00& 118& 30.0& 109& 300.0\\
4.00E-02& 5.00E+02&  221& 0.30& 189& 3.00& 167& 30.0& 156& 300.0\\
4.00E-02& 1.00E+03&  356& 0.30& 247& 3.00& 223& 30.0& 189& 300.0\\
4.00E-02& 2.00E+03&  427& 0.30& 302& 3.00& 289& 30.0& 208& 300.0\\
4.00E-02& 5.00E+03&  518& 0.30& 429& 3.00& 347& 30.0& 229& 300.0\\
4.00E-02& 1.00E+04&  672& 0.30& 549& 3.00& 435& 30.0& 245& 300.0\\
4.00E-02& 2.00E+04&  738& 0.30& 628& 3.00& 518& 30.0& 267& 300.0\\
4.00E-02& 5.00E+04&  853& 0.30& 715& 3.00& 623& 30.0& 299& 300.0\\
4.00E-02& 1.00E+05&  973& 0.30& 825& 3.00& 713& 30.0& 343& 300.0\\
4.00E-02& 2.00E+05&  1093& 0.30& 936& 3.00& 784& 30.0& 384& 300.0\\
4.00E-02& 5.00E+05&  1234& 0.30& 1024& 3.00& 825& 30.0& 426& 300.0\\
4.00E-02& 1.00E+06&  1959& 0.30& 1139& 3.00& 889& 30.0& 487& 300.0\\
\hline
\end{tabular}
\newpage
\begin{tabular}{c c c c c c c c c c}
\hline
7.00E-02& 1.00E-02&  0.19& 0.30& 0.12& 3.00& 0.11& 30.0& 0.11& 300.0\\
7.00E-02& 2.00E-02&  0.30& 0.30& 0.23& 3.00& 0.20& 30.0& 0.19& 300.0\\
7.00E-02& 5.00E-02&  0.43& 0.30& 0.37& 3.00& 0.34& 30.0& 0.32& 300.0\\
7.00E-02& 1.00E-01&  0.66& 0.30& 0.55& 3.00& 0.46& 30.0& 0.45& 300.0\\
7.00E-02& 2.00E-01&  1.30& 0.30& 0.91& 3.00& 0.83& 30.0& 0.80& 300.0\\
7.00E-02& 5.00E-01&  1.95& 0.30& 1.60& 3.00& 1.38& 30.0& 1.30& 300.0\\
7.00E-02& 1.00E-00&  4.80& 0.30& 3.12& 3.00& 2.83& 30.0& 2.67& 300.0\\
7.00E-02& 2.00E-00&  7.43& 0.30& 5.63& 3.00& 4.72& 30.0& 4.54& 300.0\\
7.00E-02& 5.00E-00&  14.2& 0.30& 10.0& 3.00& 8.30& 30.0& 8.01& 300.0\\
7.00E-02& 1.00E+01&  21.2& 0.30& 17.9& 3.00& 15.7& 30.0& 14.3& 300.0\\
7.00E-02& 2.00E+01&  38.8& 0.30& 33.4& 3.00& 28.1& 30.0& 24.8& 300.0\\
7.00E-02& 5.00E+01&  84.6& 0.30& 74.2& 3.00& 68.3& 30.0& 61.3& 300.0\\
7.00E-02& 1.00E+02&  119& 0.30& 106& 3.00& 91.3& 30.0& 84.3& 300.0\\
7.00E-02& 2.00E+02&  155& 0.30& 131& 3.00& 105& 30.0& 98.5& 300.0\\
7.00E-02& 5.00E+02&  211& 0.30& 171& 3.00& 154& 30.0& 140& 300.0\\
7.00E-02& 1.00E+03&  345& 0.30& 233& 3.00& 206& 30.0& 168& 300.0\\
7.00E-02& 2.00E+03&  413& 0.30& 291& 3.00& 270& 30.0& 191& 300.0\\
7.00E-02& 5.00E+03&  505& 0.30& 402& 3.00& 325& 30.0& 202& 300.0\\
7.00E-02& 1.00E+04&  661& 0.30& 519& 3.00& 409& 30.0& 223& 300.0\\
7.00E-02& 2.00E+04&  722& 0.30& 601& 3.00& 489& 30.0& 241& 300.0\\
7.00E-02& 5.00E+04&  840& 0.30& 691& 3.00& 600& 30.0& 247& 300.0\\
7.00E-02& 1.00E+05&  951& 0.30& 799& 3.00& 682& 30.0& 302& 300.0\\
7.00E-02& 2.00E+05&  1033& 0.30& 888& 3.00& 741& 30.0& 325& 300.0\\
7.00E-02& 5.00E+05&  1147& 0.30& 936& 3.00& 779& 30.0& 379& 300.0\\
7.00E-02& 1.00E+06&  1822& 0.30& 1041& 3.00& 821& 30.0& 411& 300.0\\
\hline
1.00E-01& 1.00E-02&  0.18& 0.30& 0.12& 3.00& 0.13& 30.0& 0.13& 300.0\\
1.00E-01& 2.00E-02&  0.28& 0.30& 0.22& 3.00& 0.22& 30.0& 0.21& 300.0\\
1.00E-01& 5.00E-02&  0.41& 0.30& 0.35& 3.00& 0.34& 30.0& 0.32& 300.0\\
1.00E-01& 1.00E-01&  0.62& 0.30& 0.51& 3.00& 0.44& 30.0& 0.45& 300.0\\
1.00E-01& 2.00E-01&  1.25& 0.30& 0.87& 3.00& 0.77& 30.0& 0.78& 300.0\\
1.00E-01& 5.00E-01&  1.87& 0.30& 1.51& 3.00& 1.23& 30.0& 1.25& 300.0\\
1.00E-01& 1.00E-00&  4.69& 0.30& 2.97& 3.00& 2.62& 30.0& 2.53& 300.0\\
1.00E-01& 2.00E-00&  7.31& 0.30& 5.43& 3.00& 4.47& 30.0& 4.38& 300.0\\
1.00E-01& 5.00E-00&  14.0& 0.30& 9.51& 3.00& 7.89& 30.0& 7.71& 300.0\\
1.00E-01& 1.00E+01&  20.3& 0.30& 16.8& 3.00& 13.7& 30.0& 13.2& 300.0\\
1.00E-01& 2.00E+01&  37.1& 0.30& 32.0& 3.00& 24.1& 30.0& 20.1& 300.0\\
1.00E-01& 5.00E+01&  82.8& 0.30& 72.5& 3.00& 64.2& 30.0& 52.8& 300.0\\
1.00E-01& 1.00E+02&  108& 0.30& 97.4& 3.00& 80.1& 30.0& 70.2& 300.0\\
1.00E-01& 2.00E+02&  143& 0.30& 106& 3.00& 91.2& 30.0& 84.2& 300.0\\
1.00E-01& 5.00E+02&  200& 0.30& 153& 3.00& 120& 30.0& 115& 300.0\\
1.00E-01& 1.00E+03&  321& 0.30& 201& 3.00& 153& 30.0& 142& 300.0\\
1.00E-01& 2.00E+03&  398& 0.30& 242& 3.00& 198& 30.0& 165& 300.0\\
1.00E-01& 5.00E+03&  468& 0.30& 337& 3.00& 251& 30.0& 189& 300.0\\
1.00E-01& 1.00E+04&  587& 0.30& 428& 3.00& 301& 30.0& 201& 300.0\\
1.00E-01& 2.00E+04&  659& 0.30& 506& 3.00& 389& 30.0& 233& 300.0\\
1.00E-01& 5.00E+04&  726& 0.30& 596& 3.00& 462& 30.0& 250& 300.0\\
1.00E-01& 1.00E+05&  812& 0.30& 684& 3.00& 538& 30.0& 289& 300.0\\
1.00E-01& 2.00E+05&  915& 0.30& 792& 3.00& 597& 30.0& 315& 300.0\\
1.00E-01& 5.00E+05&  1046& 0.30& 837& 3.00& 628& 30.0& 348& 300.0\\
1.00E-01& 1.00E+06&  1247& 0.30& 904& 3.00& 706& 30.0& 399& 300.0\\
\hline
\end{tabular}
\newpage
\begin{tabular}{c c c c c c c c c c}
\hline
2.00E-01& 1.00E-02&  0.19& 0.30& 0.13& 3.00& 0.11& 30.0& 0.11& 300.0\\
2.00E-01& 2.00E-02&  0.29& 0.30& 0.23& 3.00& 0.20& 30.0& 0.19& 300.0\\
2.00E-01& 5.00E-02&  0.42& 0.30& 0.35& 3.00& 0.33& 30.0& 0.31& 300.0\\
2.00E-01& 1.00E-01&  0.62& 0.30& 0.51& 3.00& 0.43& 30.0& 0.44& 300.0\\
2.00E-01& 2.00E-01&  1.24& 0.30& 0.86& 3.00& 0.77& 30.0& 0.77& 300.0\\
2.00E-01& 5.00E-01&  1.86& 0.30& 1.48& 3.00& 1.23& 30.0& 1.25& 300.0\\
2.00E-01& 1.00E-00&  4.65& 0.30& 2.84& 3.00& 2.65& 30.0& 2.54& 300.0\\
2.00E-01& 2.00E-00&  7.24& 0.30& 5.25& 3.00& 4.52& 30.0& 4.41& 300.0\\
2.00E-01& 5.00E-00&  13.5& 0.30& 9.08& 3.00& 8.01& 30.0& 7.82& 300.0\\
2.00E-01& 1.00E+01&  29.1& 0.30& 15.3& 3.00& 14.9& 30.0& 13.7& 300.0\\
2.00E-01& 2.00E+01&  35.0& 0.30& 28.8& 3.00& 26.8& 30.0& 20.9& 300.0\\
2.00E-01& 5.00E+01&  77.3& 0.30& 67.9& 3.00& 64.2& 30.0& 55.2& 300.0\\
2.00E-01& 1.00E+02&  100& 0.30& 90.1& 3.00& 86.3& 30.0& 75.3& 300.0\\
2.00E-01& 2.00E+02&  131& 0.30& 98.7& 3.00& 99.6& 30.0& 89.2& 300.0\\
2.00E-01& 5.00E+02&  182& 0.30& 134& 3.00& 139& 30.0& 124& 300.0\\
2.00E-01& 1.00E+03&  293& 0.30& 174& 3.00& 169& 30.0& 148& 300.0\\
2.00E-01& 2.00E+03&  337& 0.30& 216& 3.00& 211& 30.0& 167& 300.0\\
2.00E-01& 5.00E+03&  438& 0.30& 313& 3.00& 274& 30.0& 187& 300.0\\
2.00E-01& 1.00E+04&  546& 0.30& 401& 3.00& 338& 30.0& 201& 300.0\\
2.00E-01& 2.00E+04&  614& 0.30& 479& 3.00& 418& 30.0& 213& 300.0\\
2.00E-01& 5.00E+04&  672& 0.30& 527& 3.00& 493& 30.0& 220& 300.0\\
2.00E-01& 1.00E+05&  738& 0.30& 618& 3.00& 579& 30.0& 279& 300.0\\
2.00E-01& 2.00E+05&  857& 0.30& 706& 3.00& 628& 30.0& 305& 300.0\\
2.00E-01& 5.00E+05&  927& 0.30& 773& 3.00& 681& 30.0& 338& 300.0\\
2.00E-01& 1.00E+06&  1099& 0.30& 837& 3.00& 739& 30.0& 389& 300.0\\
\hline
4.00E-01& 1.00E-02&  0.20& 0.30& 0.15& 3.00& 0.14& 30.0& 0.14& 300.0\\
4.00E-01& 2.00E-02&  0.30& 0.30& 0.24& 3.00& 0.23& 30.0& 0.22& 300.0\\
4.00E-01& 5.00E-02&  0.44& 0.30& 0.36& 3.00& 0.35& 30.0& 0.33& 300.0\\
4.00E-01& 1.00E-01&  0.64& 0.30& 0.51& 3.00& 0.45& 30.0& 0.46& 300.0\\
4.00E-01& 2.00E-01&  1.25& 0.30& 0.86& 3.00& 0.78& 30.0& 0.78& 300.0\\
4.00E-01& 5.00E-01&  1.87& 0.30& 1.47& 3.00& 1.23& 30.0& 1.26& 300.0\\
4.00E-01& 1.00E-00&  4.65& 0.30& 2.82& 3.00& 2.65& 30.0& 2.54& 300.0\\
4.00E-01& 2.00E-00&  7.23& 0.30& 5.21& 3.00& 4.48& 30.0& 4.40& 300.0\\
4.00E-01& 5.00E-00&  13.4& 0.30& 9.01& 3.00& 7.91& 30.0& 7.76& 300.0\\
4.00E-01& 1.00E+01&  29.0& 0.30& 15.2& 3.00& 14.7& 30.0& 13.3& 300.0\\
4.00E-01& 2.00E+01&  34.2& 0.30& 28.6& 3.00& 26.2& 30.0& 20.1& 300.0\\
4.00E-01& 5.00E+01&  76.7& 0.30& 67.2& 3.00& 63.5& 30.0& 53.7& 300.0\\
4.00E-01& 1.00E+02&  98.6& 0.30& 89.0& 3.00& 85.1& 30.0& 74.2& 300.0\\
4.00E-01& 2.00E+02&  127& 0.30& 94.2& 3.00& 98.3& 30.0& 87.1& 300.0\\
4.00E-01& 5.00E+02&  177& 0.30& 128& 3.00& 129& 30.0& 119& 300.0\\
4.00E-01& 1.00E+03&  288& 0.30& 168& 3.00& 160& 30.0& 143& 300.0\\
4.00E-01& 2.00E+03&  330& 0.30& 209& 3.00& 198& 30.0& 154& 300.0\\
4.00E-01& 5.00E+03&  426& 0.30& 301& 3.00& 260& 30.0& 176& 300.0\\
4.00E-01& 1.00E+04&  527& 0.30& 389& 3.00& 321& 30.0& 189& 300.0\\
4.00E-01& 2.00E+04&  597& 0.30& 463& 3.00& 402& 30.0& 201& 300.0\\
4.00E-01& 5.00E+04&  642& 0.30& 516& 3.00& 476& 30.0& 210& 300.0\\
4.00E-01& 1.00E+05&  711& 0.30& 605& 3.00& 561& 30.0& 263& 300.0\\
4.00E-01& 2.00E+05&  831& 0.30& 691& 3.00& 605& 30.0& 292& 300.0\\
4.00E-01& 5.00E+05&  889& 0.30& 758& 3.00& 667& 30.0& 321& 300.0\\
4.00E-01& 1.00E+06&  1010& 0.30& 819& 3.00& 724& 30.0& 376& 300.0\\
\hline
\end{tabular}
\newpage
\begin{tabular}{c c c c c c c c c c}
\hline
7.00E-01& 1.00E-02&  0.23& 0.30& 0.16& 3.00& 0.15& 30.0& 0.15& 300.0\\
7.00E-01& 2.00E-02&  0.33& 0.30& 0.26& 3.00& 0.24& 30.0& 0.24& 300.0\\
7.00E-01& 5.00E-02&  0.47& 0.30& 0.38& 3.00& 0.37& 30.0& 0.35& 300.0\\
7.00E-01& 1.00E-01&  0.66& 0.30& 0.53& 3.00& 0.46& 30.0& 0.47& 300.0\\
7.00E-01& 2.00E-01&  1.28& 0.30& 0.88& 3.00& 0.79& 30.0& 0.79& 300.0\\
7.00E-01& 5.00E-01&  1.89& 0.30& 1.49& 3.00& 1.24& 30.0& 1.26& 300.0\\
7.00E-01& 1.00E-00&  4.70& 0.30& 2.84& 3.00& 2.66& 30.0& 2.55& 300.0\\
7.00E-01& 2.00E-00&  7.25& 0.30& 5.25& 3.00& 4.52& 30.0& 4.41& 300.0\\
7.00E-01& 5.00E-00&  13.4& 0.30& 9.00& 3.00& 8.01& 30.0& 7.82& 300.0\\
7.00E-01& 1.00E+01&  29.0& 0.30& 15.2& 3.00& 14.7& 30.0& 13.4& 300.0\\
7.00E-01& 2.00E+01&  34.1& 0.30& 28.2& 3.00& 26.1& 30.0& 20.1& 300.0\\
7.00E-01& 5.00E+01&  76.6& 0.30& 67.0& 3.00& 62.1& 30.0& 53.9& 300.0\\
7.00E-01& 1.00E+02&  98.1& 0.30& 88.1& 3.00& 84.0& 30.0& 72.4& 300.0\\
7.00E-01& 2.00E+02&  123& 0.30& 96.1& 3.00& 93.3& 30.0& 85.1& 300.0\\
7.00E-01& 5.00E+02&  173& 0.30& 128& 3.00& 122& 30.0& 118& 300.0\\
7.00E-01& 1.00E+03&  281& 0.30& 163& 3.00& 152& 30.0& 139& 300.0\\
7.00E-01& 2.00E+03&  322& 0.30& 207& 3.00& 198& 30.0& 154& 300.0\\
7.00E-01& 5.00E+03&  418& 0.30& 302& 3.00& 259& 30.0& 163& 300.0\\
7.00E-01& 1.00E+04&  519& 0.30& 390& 3.00& 320& 30.0& 189& 300.0\\
7.00E-01& 2.00E+04&  584& 0.30& 460& 3.00& 399& 30.0& 195& 300.0\\
7.00E-01& 5.00E+04&  631& 0.30& 509& 3.00& 471& 30.0& 203& 300.0\\
7.00E-01& 1.00E+05&  700& 0.30& 604& 3.00& 560& 30.0& 261& 300.0\\
7.00E-01& 2.00E+05&  819& 0.30& 691& 3.00& 604& 30.0& 289& 300.0\\
7.00E-01& 5.00E+05&  877& 0.30& 750& 3.00& 662& 30.0& 320& 300.0\\
7.00E-01& 1.00E+06&  996& 0.30& 819& 3.00& 719& 30.0& 358& 300.0\\
\hline
1.00E-00& 1.00E-02&  0.26& 0.30& 0.19& 3.00& 0.18& 30.0& 0.17& 300.0\\
1.00E-00& 2.00E-02&  0.36& 0.30& 0.29& 3.00& 0.27& 30.0& 0.26& 300.0\\
1.00E-00& 5.00E-02&  0.48& 0.30& 0.40& 3.00& 0.38& 30.0& 0.37& 300.0\\
1.00E-00& 1.00E-01&  0.69& 0.30& 0.54& 3.00& 0.47& 30.0& 0.45& 300.0\\
1.00E-00& 2.00E-01&  1.29& 0.30& 0.89& 3.00& 0.79& 30.0& 0.72& 300.0\\
1.00E-00& 5.00E-01&  1.89& 0.30& 1.49& 3.00& 1.22& 30.0& 1.18& 300.0\\
1.00E-00& 1.00E-00&  4.69& 0.30& 2.83& 3.00& 2.60& 30.0& 2.48& 300.0\\
1.00E-00& 2.00E-00&  7.21& 0.30& 5.18& 3.00& 4.41& 30.0& 4.34& 300.0\\
1.00E-00& 5.00E-00&  13.2& 0.30& 8.89& 3.00& 7.87& 30.0& 7.71& 300.0\\
1.00E-00& 1.00E+01&  28.2& 0.30& 13.9& 3.00& 14.0& 30.0& 12.4& 300.0\\
1.00E-00& 2.00E+01&  33.0& 0.30& 26.1& 3.00& 24.5& 30.0& 18.8& 300.0\\
1.00E-00& 5.00E+01&  73.2& 0.30& 65.1& 3.00& 59.9& 30.0& 50.1& 300.0\\
1.00E-00& 1.00E+02&  96.1& 0.30& 85.3& 3.00& 81.6& 30.0& 69.3& 300.0\\
1.00E-00& 2.00E+02&  118& 0.30& 93.7& 3.00& 90.7& 30.0& 82.7& 300.0\\
1.00E-00& 5.00E+02&  165& 0.30& 120& 3.00& 115& 30.0& 108& 300.0\\
1.00E-00& 1.00E+03&  263& 0.30& 151& 3.00& 146& 30.0& 121& 300.0\\
1.00E-00& 2.00E+03&  306& 0.30& 198& 3.00& 190& 30.0& 131& 300.0\\
1.00E-00& 5.00E+03&  401& 0.30& 290& 3.00& 244& 30.0& 148& 300.0\\
1.00E-00& 1.00E+04&  502& 0.30& 371& 3.00& 303& 30.0& 170& 300.0\\
1.00E-00& 2.00E+04&  563& 0.30& 441& 3.00& 380& 30.0& 181& 300.0\\
1.00E-00& 5.00E+04&  610& 0.30& 490& 3.00& 458& 30.0& 188& 300.0\\
1.00E-00& 1.00E+05&  685& 0.30& 588& 3.00& 537& 30.0& 247& 300.0\\
1.00E-00& 2.00E+05&  800& 0.30& 667& 3.00& 588& 30.0& 262& 300.0\\
1.00E-00& 5.00E+05&  856& 0.30& 729& 3.00& 649& 30.0& 292& 300.0\\
1.00E-00& 1.00E+06&  980& 0.30& 802& 3.00& 700& 30.0& 323& 300.0\\
\hline
\end{tabular}
\newpage
\begin{tabular}{c c c c c c c c c c}
\hline
2.00E-00& 1.00E-02&  0.32& 0.30& 0.25& 3.00& 0.24& 30.0& 0.22& 300.0\\
2.00E-00& 2.00E-02&  0.39& 0.30& 0.34& 3.00& 0.32& 30.0& 0.30& 300.0\\
2.00E-00& 5.00E-02&  0.53& 0.30& 0.44& 3.00& 0.42& 30.0& 0.40& 300.0\\
2.00E-00& 1.00E-01&  0.72& 0.30& 0.58& 3.00& 0.51& 30.0& 0.48& 300.0\\
2.00E-00& 2.00E-01&  1.31& 0.30& 0.92& 3.00& 0.82& 30.0& 0.74& 300.0\\
2.00E-00& 5.00E-01&  1.90& 0.30& 1.51& 3.00& 1.25& 30.0& 1.19& 300.0\\
2.00E-00& 1.00E-00&  4.70& 0.30& 2.84& 3.00& 2.62& 30.0& 2.48& 300.0\\
2.00E-00& 2.00E-00&  7.21& 0.30& 5.18& 3.00& 4.42& 30.0& 4.34& 300.0\\
2.00E-00& 5.00E-00&  13.2& 0.30& 8.89& 3.00& 7.87& 30.0& 7.61& 300.0\\
2.00E-00& 1.00E+01&  28.1& 0.30& 13.8& 3.00& 13.1& 30.0& 12.0& 300.0\\
2.00E-00& 2.00E+01&  32.0& 0.30& 25.9& 3.00& 24.0& 30.0& 17.1& 300.0\\
2.00E-00& 5.00E+01&  70.1& 0.30& 63.2& 3.00& 57.1& 30.0& 47.9& 300.0\\
2.00E-00& 1.00E+02&  93.1& 0.30& 82.5& 3.00& 77.9& 30.0& 66.1& 300.0\\
2.00E-00& 2.00E+02&  113& 0.30& 90.2& 3.00& 87.1& 30.0& 79.2& 300.0\\
2.00E-00& 5.00E+02&  158& 0.30& 115& 3.00& 109& 30.0& 101& 300.0\\
2.00E-00& 1.00E+03&  257& 0.30& 141& 3.00& 135& 30.0& 110& 300.0\\
2.00E-00& 2.00E+03&  298& 0.30& 187& 3.00& 173& 30.0& 120& 300.0\\
2.00E-00& 5.00E+03&  391& 0.30& 278& 3.00& 227& 30.0& 126& 300.0\\
2.00E-00& 1.00E+04&  489& 0.30& 357& 3.00& 289& 30.0& 156& 300.0\\
2.00E-00& 2.00E+04&  550& 0.30& 426& 3.00& 361& 30.0& 163& 300.0\\
2.00E-00& 5.00E+04&  588& 0.30& 470& 3.00& 435& 30.0& 173& 300.0\\
2.00E-00& 1.00E+05&  670& 0.30& 569& 3.00& 520& 30.0& 222& 300.0\\
2.00E-00& 2.00E+05&  781& 0.30& 651& 3.00& 561& 30.0& 249& 300.0\\
2.00E-00& 5.00E+05&  832& 0.30& 708& 3.00& 632& 30.0& 270& 300.0\\
2.00E-00& 1.00E+06&  958& 0.30& 789& 3.00& 688& 30.0& 301& 300.0\\
\hline
3.00E-00& 1.00E-02&  0.37& 0.30& 0.31& 3.00& 0.29& 30.0& 0.27& 300.0\\
3.00E-00& 2.00E-02&  0.44& 0.30& 0.38& 3.00& 0.34& 30.0& 0.32& 300.0\\
3.00E-00& 5.00E-02&  0.59& 0.30& 0.48& 3.00& 0.44& 30.0& 0.42& 300.0\\
3.00E-00& 1.00E-01&  0.76& 0.30& 0.61& 3.00& 0.53& 30.0& 0.50& 300.0\\
3.00E-00& 2.00E-01&  1.33& 0.30& 0.94& 3.00& 0.83& 30.0& 0.76& 300.0\\
3.00E-00& 5.00E-01&  1.92& 0.30& 1.53& 3.00& 1.26& 30.0& 1.20& 300.0\\
3.00E-00& 1.00E-00&  4.71& 0.30& 2.85& 3.00& 2.63& 30.0& 2.49& 300.0\\
3.00E-00& 2.00E-00&  7.21& 0.30& 5.18& 3.00& 4.42& 30.0& 4.34& 300.0\\
3.00E-00& 5.00E-00&  13.2& 0.30& 8.89& 3.00& 7.87& 30.0& 7.55& 300.0\\
3.00E-00& 1.00E+01&  28.0& 0.30& 13.2& 3.00& 12.3& 30.0& 11.2& 300.0\\
3.00E-00& 2.00E+01&  30.1& 0.30& 24.0& 3.00& 22.1& 30.0& 15.7& 300.0\\
3.00E-00& 5.00E+01&  67.9& 0.30& 60.1& 3.00& 55.0& 30.0& 44.1& 300.0\\
3.00E-00& 1.00E+02&  90.0& 0.30& 79.2& 3.00& 74.2& 30.0& 62.3& 300.0\\
3.00E-00& 2.00E+02&  106& 0.30& 87.1& 3.00& 83.8& 30.0& 75.7& 300.0\\
3.00E-00& 5.00E+02&  148& 0.30& 108& 3.00& 101& 30.0& 95.9& 300.0\\
3.00E-00& 1.00E+03&  241& 0.30& 130& 3.00& 123& 30.0& 102& 300.0\\
3.00E-00& 2.00E+03&  280& 0.30& 171& 3.00& 160& 30.0& 110& 300.0\\
3.00E-00& 5.00E+03&  371& 0.30& 265& 3.00& 210& 30.0& 119& 300.0\\
3.00E-00& 1.00E+04&  473& 0.30& 341& 3.00& 272& 30.0& 145& 300.0\\
3.00E-00& 2.00E+04&  533& 0.30& 407& 3.00& 349& 30.0& 155& 300.0\\
3.00E-00& 5.00E+04&  572& 0.30& 451& 3.00& 419& 30.0& 163& 300.0\\
3.00E-00& 1.00E+05&  653& 0.30& 550& 3.00& 501& 30.0& 201& 300.0\\
3.00E-00& 2.00E+05&  767& 0.30& 635& 3.00& 548& 30.0& 232& 300.0\\
3.00E-00& 5.00E+05&  818& 0.30& 689& 3.00& 611& 30.0& 252& 300.0\\
3.00E-00& 1.00E+06&  940& 0.30& 770& 3.00& 667& 30.0& 280& 300.0\\
\hline
\hline
\end{tabular}

$^a$ In $M_\odot$ yr$^{-1}$.

$^b$ The notation $aEb$ denotes $a\times 10^b$.

\newpage

\begin{figure}
\centerline{\psfig{figure=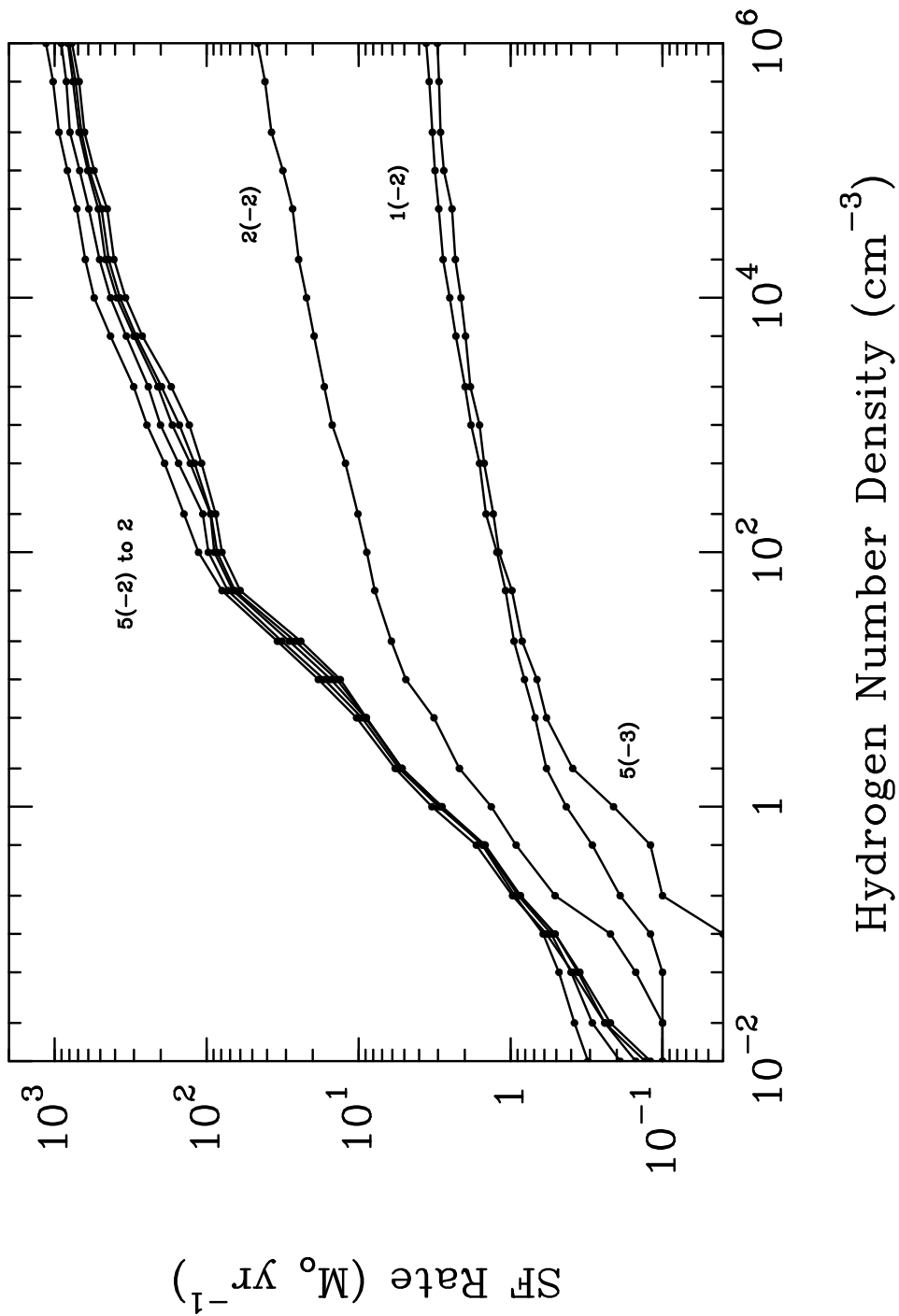,width=6in}}
\label{figure1}
\caption{ The star formation rate as a function of hydrogen number density,
for various metallicities, a total mass of $M_{\rm g}$ and a background
star formation rate of $S_{\rm b}=3$ M$_\odot$ yr$^{-1}$. The various curves
are labelled by the ambient metallicity relative to Solar.}
\end{figure}

\newpage

\begin{figure}
\centerline{\psfig{figure=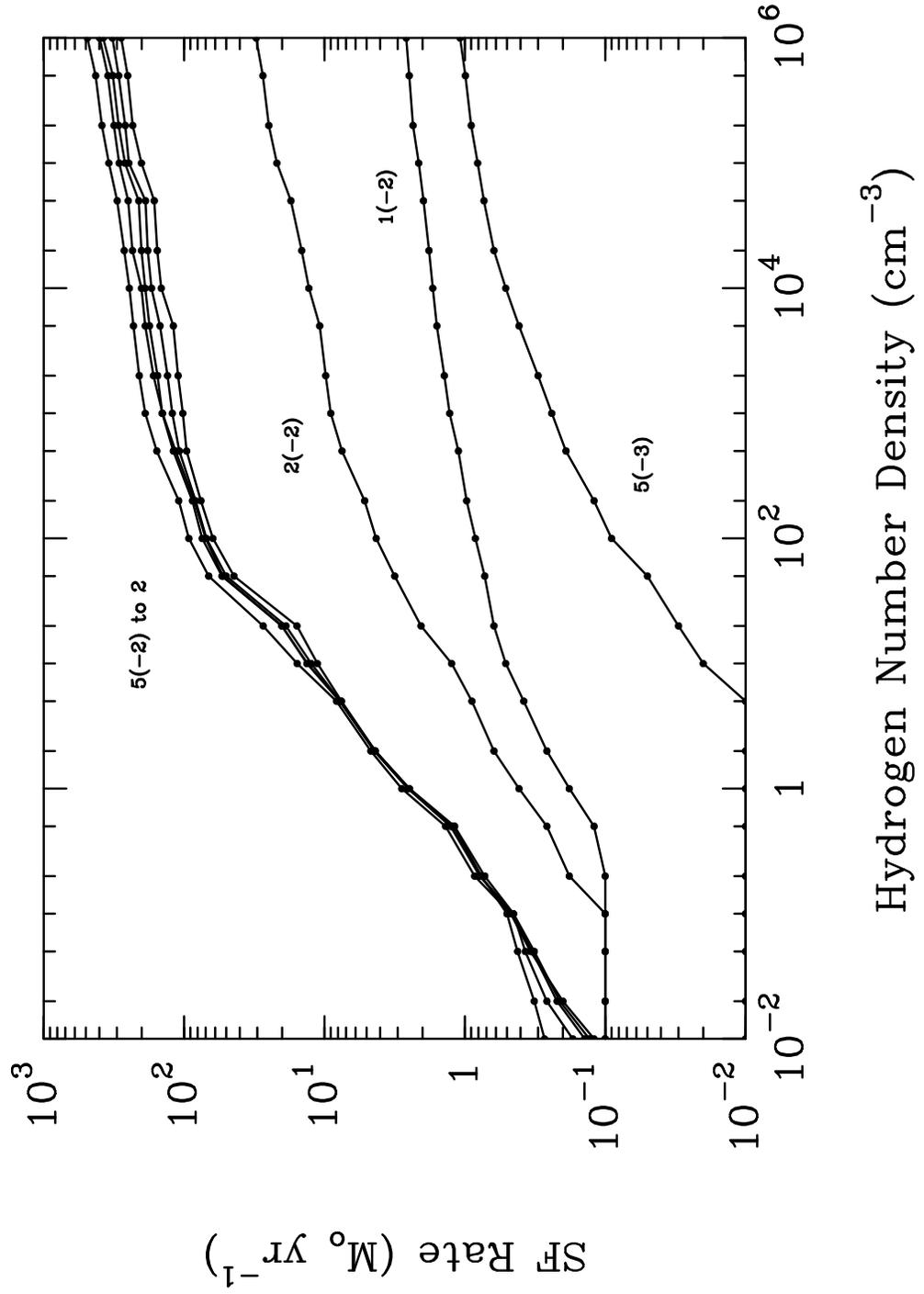,width=6in}}
\label{figure2}
\caption{Same as Figure 1, for $S_{\rm b}=300$ M$_\odot$ yr$^{-1}$.}
\end{figure}

\end{document}